\documentclass[aip,jmp,author-year,reprint,groupedaddress,nofootinbib,nobibnotes,onecolumn]{revtex4-1}

\usepackage{amsmath}
\usepackage{amssymb}
\usepackage{amsfonts}
\usepackage{dsfont}
\usepackage{graphicx}
\usepackage{color}
\usepackage{subfig}
\usepackage[latin2]{inputenc}
\usepackage{hyperref}
\usepackage{booktabs}
\usepackage{multirow}
\usepackage[]{natbib}

\usepackage{scalefnt}
\usepackage{placeins}

\newcommand{\exph}[1]{\exp\left( #1 \right)}
\newcommand{\tp}{^{\dag}}
\newcommand{\rd}{\mathrm{d}}
\newcommand{\tr}{\mathrm{tr}}
\makeatletter
\newcommand{\vast}{\bBigg@{3}}
\newcommand{\Vast}{\bBigg@{4}}
\makeatother

\begin{document}

\title{A Random Matrix Approach to Credit Risk}
\author{Michael C. M\"unnix,\footnote{Both authors contributed equally to this work.}}
\email{michael@muennix.com}
\author{Rudi Sch\"afer,$^{a)}$}
\email{rudi.schaefer@uni-duisburg-essen.de}
\author{Thomas Guhr}
\email{thomas.guhr@uni-duisburg-essen.de}
\affiliation{Faculty of Physics, University of Duisburg-Essen, Duisburg, Germany}

\begin{abstract}
We estimate generic statistical properties of a structural credit risk model by considering an ensemble of correlation matrices. This ensemble is set up by Random Matrix Theory.
We demonstrate analytically that the presence of correlations severely limits the effect of diversification in a credit portfolio if the correlations are not identically zero. The existence of correlations alters the tails of the loss distribution considerably, even if their average is zero. Under the assumption of randomly fluctuating correlations, a lower bound for the estimation of the loss distribution is provided.

\end{abstract}

\maketitle

\section{Introduction}
The financial crisis of 2008--2009 clearly revealed that an improper estimation of credit risk can lead to dramatic effects on the world's economy. The vast underestimation of risks embedded in credits for the subprime housing markets induced a chain reaction that propagated into the worldwide economy. A better estimation of credit risk (see, eg, \cite{b_bluhm02,b_bielecki05,b_duffie03,b_lando04,b_mcneil05}) is therefore of vital interest.

We can distinguish two fundamentally different approaches to credit risk modeling (see, eg, \cite{giesecke04b}): the structural and the reduced--form approach.
\emph{Structural models} have a long history, going back to the work of  \cite{black73} and \cite{metron74}. The \emph{Merton model} assumes a zero--coupon debt structure with a fixed time to maturity. The equity of the company is modeled by a stochastic process, for example describing the price of the company's stock. Thus, it can be seen as a creditor's call option on the obligor's assets. The risk of default and the associated recovery rate, the residual payment in case of a loss, are modeled by the company's equity at maturity.
\emph{Reduced--form models} attempt to capture the dependence of default and recovery rates on macroeconomic risk factors. Both quantities are modeled as independent stochastic variables.
Some well known reduced--form model approaches can be found in \cite{jarrow95,jarrow97,duffie99,hull00,schonbucher03}.
\emph{First passage models}  were first introduced by \cite{black76}. Although they are often referred to as structural models, they do, in fact, follow a mixed approach. Similar to Merton's model, the market value of a company is modeled by a stochastic process. However, in the first passage models a default occurs whenever this market value hits a certain threshold for the first time. The recovery rates are typically modeled independently, for example, by a reduced--form model, see \cite{asvanunt09a, asvanunt09b}, or are even assumed to be constant, see, eg,  \cite{giesecke04b}. Recent approaches aim at improving fist passage models by including the chance of full recovery, even if a company's market value is below the threshold, see \cite{katz10}, and estimating correlations between default probabilities of industry sectors, see \cite{rosenow09}.
Reduced--form and first passage models are implemented in commercial software solutions, for example, \emph{CreditMetrics} by JP Morgan (see \cite{CreditMetrics}), \emph{CreditPortfolioView} by McKinsey \& Company (\citeyear{CreditPortfolioView}) or \emph{CreditRisk+} by Credit Suisse (\citeyear{CreditRisk+}).
As there can be a strong connection between default risks and recovery rates, the chances of large losses are often underestimated in the reduced--form and first passage models, see \cite{SchaeferKoivusalo2011,KoivusaloSchaefer2011}. Structural models do not require this separation.

Structural models provide a ``microscopical'' tool to study credit risk as the defaults and recoveries are traced back to stochastic processes modeling the state of individual obligors.
For a portfolio of credits, such as collateralized debt obligations (CDOs), correlations represent a key factor that influences its risk.
The benefit of diversification, ie, the reduction of risk by increasing the portfolio size, is severely limited by the presence of even weak correlations.
This has been demonstrated for the case of constant positive correlations, both in the first passage model with constant recovery (\cite{Schoenbucher2001,Glasserman2003}) and in the Merton model (\cite{schaefer07,SchaeferKoivusalo2011}).
The key problem in estimating the credit risk of a realistic portfolio  is of course the huge number of parameters involved. 
This is precisely where approaches from statistical physics can be most helpful: the state of a system with many degrees of freedom is, 
under certain conditions, described by few macroscopic observables. In the thermodynamic equilibrium, these are energy, temperature, 
pressure, etc. Ergodicity holds, ie, time and ensemble average yield the same results. A somewhat similar situation exists for spectral statistics in quantum chaotic systems, see \cite{guhr98}. A moving average over one long spectrum equals an ensemble average over 
random matrices, if the number of levels is very large. 
Originally, random matrix theory was developed in the 1950s to describe the spectra of heavy nuclei, see \cite{meh04}.
Here we transfer this idea to large credit portfolios involving correlated assets. 
In the case of a great many contracts, we expect a self--averaging property which then should allow to average over an
ensemble of random correlation matrices.  We manage to carry out this approach largely analytically. 
We obtain estimates for the price and the loss distribution in which the complicated effects of all correlations are indeed reduced to a single 
parameter measuring the correlation strength.  
This paper is organized as follows. A structural credit risk model with random correlations is considered in section \ref{s:cr:model}. We demonstrate an application of this model in section \ref{s:cr:application} and conclude our findings in section \ref{s:cr:summary}.

\section{A structural credit risk model}
 \label{s:cr:model}
Our model is based on Merton's original model. Our aim is to analytically describe the impact of correlations on the losses of a credit portfolio.
A default occurs if the price $V_{k}$ of the $k$-th asset is below the face value $F_{k}$ at maturity time $T$. The size of the loss then depends on how far $V_{k}$ is below the face value $F_{k}$.
We assume that the prices in a portfolio of $K$ assets follow a geometric Brownian motion.
An overview of the model's input parameters is given in Tab. \ref{tab:cr:input}.

\begin{table}
\centering
\begin{tabular}{l l l}
\hline\hline 
Variable & Description & Unit\\
\hline 
$K$ & Number of assets & -- \\
$T$ & Time of maturity & [year]\\
$\sigma_{k}$ & volatility of the $k$-th asset & [year]$^{-1/2}$\\
$\mu_{k}$ & drift of the $k$-th asset & [year]$^{-1}$\\
$N$ & Parameter to control correlations, & --
\\&$N\to\infty$: uncorrelated limit & \\
$V_{k,0}$ & start price of the $k$-th asset & [currency]\\
$F_{k}$ & face value of the $k$-th asset & [currency]\\
\hline
\hline
\end{tabular}
\caption[Input of the structural credit risk model]{Input of the structural credit risk model.}
\label{tab:cr:input}
\end{table}

\subsection{Average price distribution}
For the sake of simplicity, let us first consider the case of a Brownian motion for the average price distribution. Later on this can be easily mapped to the geometric Brownian motion by a simple substitution. For a Brownian motion, the probability density function (pdf) of the price vector  $V$ of $K$ assets at maturity $T$ is described by 
\begin{equation}
p^{\rm (mv)}(V, \Sigma) 
=
\frac{1}{\sqrt{2\pi T}^K}
\frac{1}{\sqrt{\mathrm{det}(\Sigma)}}
\exph{-\frac{1}{2T}(V-\mu T)\tp\Sigma^{-1}(V-\mu T)} 
\label{eq:cr:gbm}
\end{equation}
Here, $\Sigma$ is the covariance matrix and $\mu$ is the drift vector.
For later convenience we can express this as a Fourier transform,
\begin{equation}
p^{\rm (mv)}(V, \Sigma) 
=
\frac{1}{(2\pi)^K}
\int \exph{-i \omega\tp (V-\mu T)}
\exph{-\frac{T}{2}\omega\tp \Sigma\omega}
\rd [\omega]\label{eq:creditrmstart}
\end{equation}
Equation (\ref{eq:cr:gbm}) gives the pdf of asset prices in the case of a correlated Brownian motion. However, we are not interested in the impact of a specific correlation matrix. Instead we want to estimate the general impact of correlations. To this end, we want to average over all possible  correlation matrices and disclose the general statistical behavior of the system. This will enable us to make a profound statistical statement.

We use a random matrix approach to calculate the average price distribution for random correlations where the average correlation level is zero. By averaging over all possible combinations of random variables, we obtain the average price distribution $\langle p^{\rm (mv)}(V) \rangle$ under these assumptions.
To achieve this we replace the covariance matrix $\Sigma$ by 
\begin{equation}
\Sigma_{W}=S W W\tp S
\end{equation}
where $S=\mathrm{diag}(\sigma_{1},\dots,\sigma_{K})$ contains the standard deviations and $W \in \mathbb{R}^{K\times N}$ is a random matrix. 
The entries of $W$ are independent and Gaussian distributed,
\begin{equation}
p^{\rm (corr)}(W) = \sqrt{\frac{N}{2\pi}}^{KN}  \exph{ - \frac{N}{2} \tr W\tp W }
\end{equation}
with variance $1/N$. This results in a correlation matrix $W W\tp$ with average correlation zero. 
With the parameter $N$ we can control how strongly the entries of $W W\tp$ fluctuate.
For $N \to \infty$, we obtain the pdf of a unit matrix. This represents the uncorrelated case. For $N\ge K$, we obtain an invertible covariance matrix with random entries. The case $N<K$ is disregarded as the resulting matrix is not invertible which is usually required for applications in risk management.
When inserting this ansatz into Eq. (\ref{eq:creditrmstart}), we obtain
\begin{eqnarray}
\langle p^{\rm (mv)}(V) \rangle&=& \int p^{\rm (corr)}(W) p^{\rm (mv)}(V, S W W\tp S) \rd [W] \\
&=& \frac{\sqrt{N}^{NK}}{(2\pi)^K}
\int \exph{-i \omega\tp V}
\frac{1}{\sqrt{\mathrm{det}(N I + T S\omega \omega\tp S)}^{N}} 
\rd [\omega]
\end{eqnarray}
where $I$ denotes the unit matrix. A detailed derivation is given in appendix \ref{app:aveprice}. 
Here we choose $\mu=0$. We will reintroduce the drift later on, when we make the substitution for the geometric Brownian motion.
The determinant can be written as
\begin{eqnarray}
\mathrm{det}(N I + T S\omega \omega\tp S) 
&= N^{K}\left(1+\frac{T}{N}\omega\tp S S\omega\right)
\end{eqnarray}
because the matrix $S\omega \omega\tp S$ has rank one.
Hence, we arrive at
\begin{equation}
\langle p^{\rm (mv)}(V) \rangle = \frac{1}{(2\pi)^{K}}
\int \exph{-i \omega\tp V}
\frac{1}{(1 + (T/N)\omega\tp S S\omega )^{N/2}}
\rd [\omega]
\end{equation}
This integral can be calculated by using the Gamma function (see \cite{b_olver74}) in the form
\begin{eqnarray}
\frac{\Gamma(x)}{a^{x}}= \int\limits_{0}^{\infty}z^{x-1}\exph{-az}\rd z\quad,\quad x>0,\ a>0 
\end{eqnarray}
We identify $a^{-x}$ with $((1 + (T/N)\omega\tp S S\omega ))^{-N/2}$ and obtain
\begin{eqnarray}
\langle p^{\rm (mv)}(V) \rangle
&=&
\frac{1}{(2\pi)^{K}}
\frac{1}{\Gamma(N/2)}
\left(
\prod\limits_{k=1}^{K}
\frac{1}{\sigma_{k}}
\right)
\nonumber\\&&\times
\int\limits_{0}^{\infty}
z^{\left(\frac{N}{2}-1\right)}
\exph{-z}
\sqrt{\frac{\pi N}{zT}}^{K}
\exph{-\frac{N}{4 Tz}\sum\limits_{k=1}^{K}\frac{V_{k}^{2}}{\sigma_{k}^{2}}}\rd z
\end{eqnarray}
as worked out in appendix \ref{app:aveprice2}. This integral is a representation of the Bessel function of the second kind $\mathcal{K}$ of the order $(K-N)/2$, see \cite{b_watson44}. Thus, we obtain 
\begin{equation} 
\langle p^{\rm (mv)}(V) \rangle=
\sqrt{\frac{N}{2\pi T}}^{K}
\frac{2^{1-\frac{N}{2}}}{\Gamma(N/2)}
\left(
\prod\limits_{k=1}^{K}
\frac{1}{\sigma_{k}}
\right)
\sqrt{\frac{N}{T}\sum\limits_{k=1}^{K}\frac{V_{k}^{2}}{\sigma_{k}^{2}}}^{\frac{N-K}{2}}
\mathcal{K}_{\frac{N-K}{2}}\left(\sqrt{\frac{N}{T} \sum\limits_{k=1}^{K} \frac{V_{k}^{2}}{\sigma_{k}^{2}}} \right)
\label{eq:cr:meanPV}
\end{equation}
for the average distribution of $p^{\rm (mv)}(V)$ if assuming a randomly distributed correlation matrix and an average correlation level of zero.
We stated earlier that we include $N$ in the distribution of the random matrices $W$ in order to render the variance of the average price distribution $N$-independent. The variances only depend on $T$ and $\sigma_{k}$, as discussed in appendix \ref{appnorm}. The parameter $N$ is only used to control the correlations.
\begin{figure}[t]
\begin{center}
\subfloat[linear scale]
{
\includegraphics[width=0.48\textwidth]{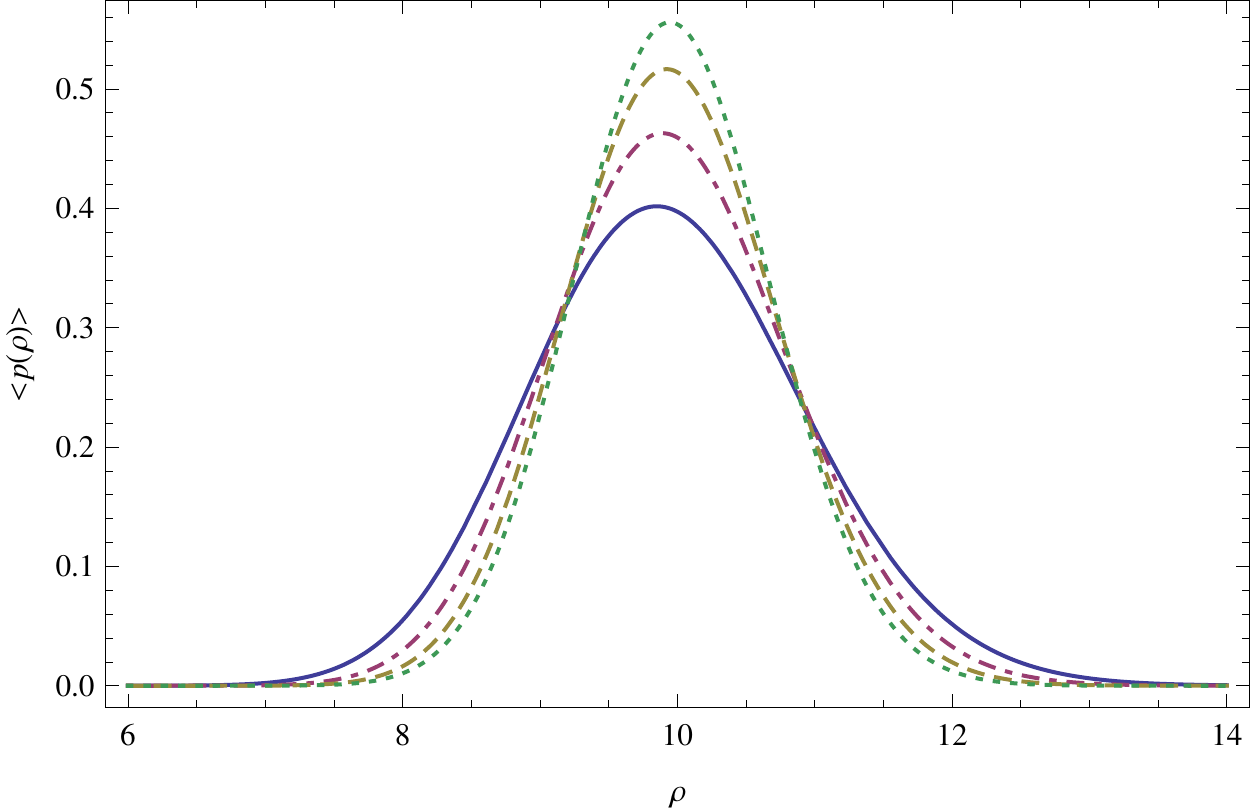}
}
\subfloat[logarithmic scale]
{
\includegraphics[width=0.48\textwidth]{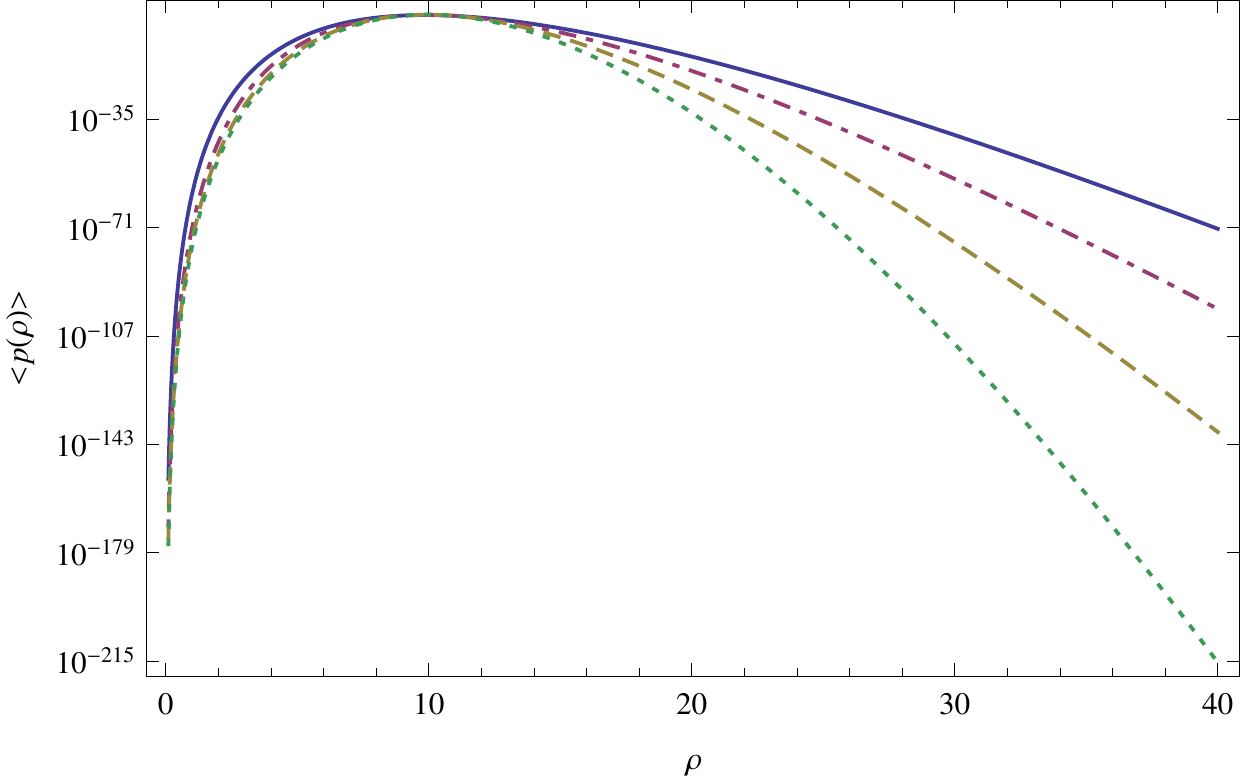}
}
\caption{Illustration of the average price distribution $\langle p^{\rm (mv)}(\rho)\rangle$  for $T=1$, $K=50$ and different values for $N$. 
Solid, dashed--dotted, dashed and dotted lines correspond to $N=K$, $2K$, $5K$ and $30K$, respectively.
}
\label{fig:cr:meanpvrho}
\end{center}
\end{figure}
In hyperspherical coordinates, Equation \ref{eq:cr:meanPV} depends only on the hyperradius  
\begin{eqnarray}
\rho \equiv \sqrt{\sum\limits_{k=1}^{K}\frac{V_{k}^{2}}{\sigma_{k}^{2}}} 
\end{eqnarray}
This leads to the expression
\begin{equation} 
\langle p^{\rm (mv)}(\rho) \rangle=
\sqrt{\frac{N}{2\pi T}}^{K}
\frac{2^{1-\frac{N}{2}}}{\Gamma(N/2)}
\rho^{\frac{N+K-1}{2}}  \sqrt{\frac{N}{T}}^{\frac{N-K}{2}}
\mathcal{K}_{\frac{N-K}{2}}\left( \rho \sqrt{\frac{N}{T}} \right) 
\label{eq:cr:meanPVrho}
\end{equation}
for the hyperradial density function, cf.~appendix \ref{appnorm}.
We illustrate this density function in Figure~\ref{fig:cr:meanpvrho} for $K=50$ and different values of $N$.

\begin{figure}[t]
\begin{center}
\subfloat[$N$=2]
{
\includegraphics[width=0.48\textwidth]{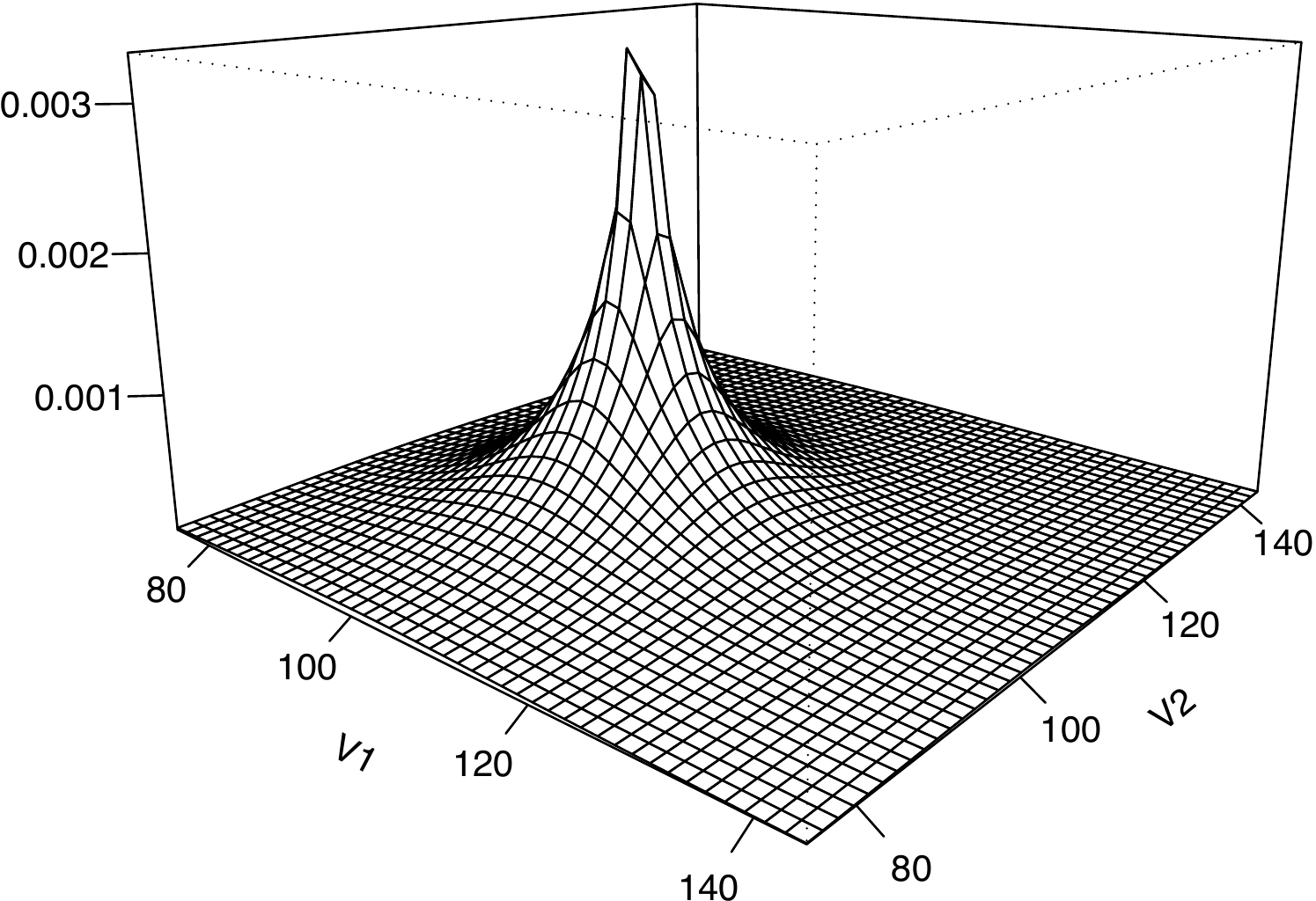}
}
\subfloat[$N$=100]
{
\includegraphics[width=0.48\textwidth]{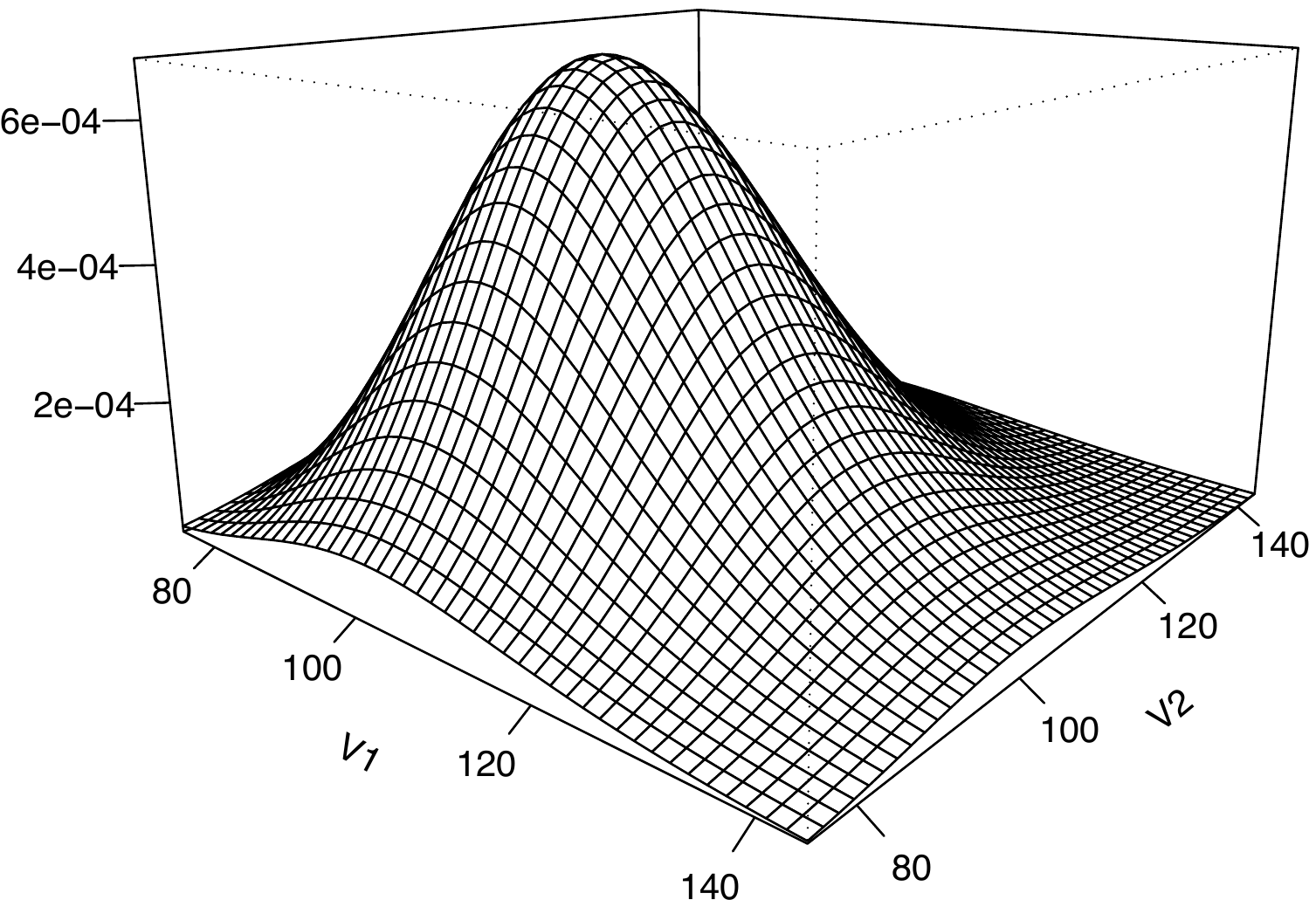}
}
\caption{Illustration of the average price distribution $\langle p^{\rm (mv)}(V)\rangle$ with a geometric Brownian motion for $T=1$, $K=2$, $V_{k,0}=100$, $\mu =0.05$ and different values for $N$. Both distributions have the identical standard deviation $\hat\sigma\approx 16$ ($\sigma=0.15$). For $N=2$, we obtain a heavy-tailed distribution while the uncorrelated limit is reached for $N=100$.}
\label{fig:cr:meanpvlog}
\end{center}
\end{figure}

We obtain the average price distribution in case of a geometric Brownian motion by a simple substitution $V_{k} \to \widehat V_{k}$,
\begin{equation} 
\langle p^{\rm (mv)}(V) \rangle = 
\sqrt{\frac{N}{2\pi T}}^{K}
\frac{2^{1-\frac{N}{2}}}{\Gamma(N/2)}
\left(
\prod\limits_{k=1}^{K}
\frac{1}{\sigma_{k}V_{k}}
\right)
\sqrt{\frac{N}{T}\sum\limits_{k=1}^{K}\frac{\widehat V_{k}^{2}}{\sigma_{k}^{2}}}^{\frac{N-K}{2}}
\mathcal{K}_{\frac{N-K}{2}}\left(\sqrt{\frac{N}{T} \sum\limits_{k=1}^{K} \frac{\widehat V_{k}^{2}}{\sigma_{k}^{2}}} \right)
\label{eq:cr:meanPVlog}
\end{equation}
with
\begin{eqnarray}
\widehat V_{k}=\ln\left(\frac{V_{k}}{V_{k,0}}\right) - \left(\mu_{k}-\frac{\sigma_{k}^{2}}{2}\right)T
\end{eqnarray}
Here, the parameter $\sigma_{k}$ refers to the standard deviation of the underlying Brownian motion, ie, the volatility of asset returns. The resulting prices thus have the variance
\begin{eqnarray}
\hat\sigma_{k}=\sqrt{\exph{\sigma_{k}^2 T + 2\mu } (\exph{\sigma_{k}^2 T}-1) V^{2}_{k,0}} 
\end{eqnarray}
where $V_{k,0}$ are the starting prices at $t=0$.
Figure \ref{fig:cr:meanpvlog} shows the distribution of prices based on a geometric Brownian motion, as given in Eq. (\ref{eq:cr:meanPVlog}). The findings are similar to the case of the Brownian motion. While we obtain a narrow but heavy-tailed distribution for $N=K$, the distribution slowly approaches an uncorrelated bivariate log-normal distribution with increasing values of $N$.

\subsection{Loss distribution}
We now turn to the calculation of the loss distribution.
A default occurs if the price $V_{k}$ at maturity $T$, corresponding to the obligor's equity, is below the face value $F_{k}$. The size of the loss is given by the difference of $F_{k}$ and $V_{k}$. Even if a loss occurs, the creditor might not lose all money that he lent, because the obligor is still able to pay back the amount $V_{k}$. In order to compare losses in a portfolio of credits, we have to normalize them by the corresponding face value. We define the normalized loss $L_{k}$ of the $k$-th asset as
\begin{equation}
L_{k}=
\left\{ \begin{array}{cll}
\frac{F_{k}-V_{k}}{F_{k}} &,\quad  V_{k}<F_{k} & \mathrm{(default)}\\
0 &,\quad \mathrm{else} & \mathrm{(no\ default)}
\end{array}
\right.
\label{eq:cr:lossdef}
\end{equation}
We observe that the prices have to be positive in Eq.~(\ref{eq:cr:lossdef}). Therefore we assume in all further considerations that the underlying process of the price distribution follows a geometric Brownian motion.

When calculating the overall loss of a portfolio, we have to weight each loss by its face value in relation to the sum of all portfolio face values,
\begin{equation}
L = \sum\limits_{k=1}^{K}f_{k}L_{k} \quad , \quad f_{k} = \frac{F_{k}}{\sum_{i=1}^{K} F_{i}}
\end{equation}
We integrate over the pdf of prices and filter for those that lead to a given total loss $L$. By the above stated definitions, we can  define a filter for the total loss at maturity time $T$. In the next step we express the filter using a Fourier transformation. Eventually, we separate those terms that correspond to a default and those that describe the price above the face value $F_{k}$. 
\begin{eqnarray}
p^{\rm (loss)}(L) &=&  \int\limits_{0}^{\infty} \rd [V]p^{\rm (mv)}(V)\delta\left(L-\sum\limits_{k=1}^{K}f_{k}L_{k}\right)\\
&=& \int\limits_{0}^{\infty} \rd [V]p^{\rm (mv)}(V)
\frac{1}{2\pi}\int\limits_{-\infty}^{+\infty} \rd\nu \exph{-i\nu L + i\nu \sum\limits_{k=1}^{K}f_{k}L_{k}}\\
&=& \frac{1}{2\pi}\int\limits_{-\infty}^{+\infty} \rd\nu \exph{-i\nu L}
\int\limits_{0}^{\infty} \rd [V]\exph{i\nu \sum\limits_{k=1}^{K}f_{k}L_{k}} p^{\rm (mv)}(V)\\
&=& \frac{1}{2\pi}\int\limits_{-\infty}^{+\infty} \rd\nu \exph{-i\nu L}
\nonumber\\&&\times
\prod_{k=1}^{K}
\left[
\int\limits_{0}^{F_{k}} \rd V_{k}\exph{i\nu f_{k} \left(1-\frac{V_{k}}{F_{k}}\right)}
+
\int\limits_{F_{k}}^{\infty} \rd V_{k}
\right]
p^{\rm (mv)}(V)
\label{eq:cr:loss:split}
\end{eqnarray}
Here, the expression in the square brackets acts as an operator, because $p^{\rm (mv)}(V)$ does not necessarily factorize. We will use this ansatz to calculate the average loss distribution in the next section. However, Eq. (\ref{eq:cr:loss:split}) can be used to calculate the loss distribution if the actual price distribution is known, ie, the statistical dependence and the underlying process are estimated. To  prepare for this, it is handy to write Eq. (\ref{eq:cr:loss:split}) as a combinatorial sum,
\begin{eqnarray} 
p^{\rm (loss)}(L)
&=& \frac{1}{2\pi}\int\limits_{-\infty}^{+\infty} \rd\nu \exph{-i\nu L}
\\ &&\times
\sum_{k=1}^{K}
\sum_{j=1}^{K \choose k}
\left[
\prod\limits_{l \in \mathrm{Perm}(j,k,K) \atop}
\int\limits_{0}^{F_{l}} \rd V_{l}
\exph{i\nu f_{l}\frac{F_l-V_{l}}{F_{l}}}
\prod\limits_{q \in \{1 \dots K \} \atop\backslash \mathrm{Perm}(j,k,K)}
\int\limits_{F_{q}}^{\infty} \rd V_{q}
\right]
p^{\rm (mv)}(V)
\nonumber
\label{eq:cr:combsum}
\end{eqnarray}
where $\mathrm{Perm}(j,k,K)$ is the $j$-th permutation of $k$ elements of the set $\{1\dots K\}$. For example, if $K=3$ and $k=2$, we obtain, $\mathrm{Perm}(1,2,3)= \{1,2\}$, $\mathrm{Perm}(2,2,3)= \{2,3\}$ and $\mathrm{Perm}(3,2,3)= \{1,3\}$.
However, Eq. (\ref{eq:cr:combsum}) might need to be estimated numerically, depending on the complexity of the price distribution $p^{\rm (mv)}(V)$. In section \ref{eq:cr:homoport}, we will simplify this combinatorial sum for a homogeneous portfolio and the average price distribution $\langle p^{\rm (mv)}(V)\rangle$.

\subsection{Average loss distribution}
Now we have developed all necessary tools to model the average distribution of losses, under the assumption of random correlations and an average correlation level of zero.
We start by inserting the average price distribution in a component-wise notation (see appendix \ref{app:aveprice2}) into the loss distribution (\ref{eq:cr:loss:split}),
\begin{equation}
\left\langle p^{\rm (loss)}(L)\right\rangle
= 
\frac{1}{2\pi\Gamma(N/2)}
\int\limits_{0}^{\infty}
\rd z\ 
z^{\left(\frac{N}{2}-1\right)}
\exph{-z}
\int\limits_{-\infty}^{+\infty} \rd\nu \exph{-i\nu L}
r(\nu,z)
\label{eq:cr:wurst}
\end{equation}
with
\begin{eqnarray}
r(\nu,z) &=&
\prod_{k=1}^{K}
\left[
\int\limits_{0}^{F_{k}} \rd V_{k}
\exph{i\nu f_{k}\frac{F_k-V_{k}}{F_{k}}}
+
\int\limits_{F_{k}}^{\infty} \rd V_{k}
\right]
\nonumber\\&&\times
\frac{\sqrt{N}}{2\sigma_{k}V_{k}\sqrt{\pi z T}}
\exph{-\frac{N(\ln(V_{k}/V_{k,0})-(\mu-\sigma^{2}/2)T) ^{2}}{4zT\sigma_{k}^{2}}}
\end{eqnarray}
We carry out a second order approximation of this expression in appendix \ref{app:secorder} and arrive at
\begin{eqnarray} 
\left\langle p^{\rm (loss)}(L)\right\rangle= 
\frac{1}{\sqrt{2\pi}\Gamma(N/2)}
\int\limits_{0}^{\infty}
\rd z\ 
z^{\left(\frac{N}{2}-1\right)}
\exph{-z}
\frac{1}{\sqrt{\widehat m_{2}(z)}} \exph{-\frac{(L-\widehat m_{1}(z))^2}{2\widehat m_{2}(z)}}
\label{eq:cr:simpelapprox}
\end{eqnarray}
with
\begin{align}
\widehat m_{1}(z)&=\sum\limits_{k=1}^{K}f_{k} m_{1,k}(z)\\
\widehat m_{2}(z)&=\sum\limits_{k=1}^{K}f_{k}^{2}(m_{2,k}(z)-m_{1,k}(z)^2)
\end{align}
and
\begin{eqnarray}
m_{j,k}(z)&=&
\frac{\sqrt{N}}{2\sigma_{k}\sqrt{\pi z T}}
\int\limits_{0}^{F_{k}}
\frac{1}{V_{k}}
\left(\frac{F_k-V_{k}}{F_{k}}\right)^{j}
\nonumber\\&&\times
\exph{-\frac{N(\ln(V_{k}/V_{k,0})-(\mu-\sigma^{2}/2)T) ^{2}}{4zT\sigma_{k}^{2}}}
\rd V_{k}
\end{eqnarray}
However, the convergence radius of the power series expansion involved in this approximation is one. Although we consider large portfolios $K$, ie, $f_{k}$ is small, $\nu$ runs from $-\infty$ to $+\infty$. This second-order approximation might describe the default terms adequately. However, the non-default terms, corresponding to a delta peak at $L=0$ require $\nu$ to run from $-\infty$ to $+\infty$. Thus, the non-default terms cannot be approximated using this second-order approximation. To circumvent this problem we develop an improved approximation in section \ref{eq:cr:homoport}.

Due to the complexity of $\hat m_{1}(z)$ and $\hat m_{2}(z)$, the $z$ integral needs to be evaluated numerically.
We present this for the example of a homogeneous portfolio.

\subsection{Homogeneous portfolios}
\label{eq:cr:homoport}
In case of a homogeneous portfolio, in which all credits have the same face value $F_k=F$ and the same variance $\sigma_k^{2}=\sigma^{2}$ and initial value $V_{k,0}=V_0$, the weights can be simplified to
\begin{equation}
f_{k} = \frac{1}{K}
\end{equation}
As $m_{1,k}(z)$ and $m_{1,k}(z)$ become identical for every $k$, we denote them by $m_{1}(z)$ and $m_{1}(z)$ leading to
\begin{eqnarray}
\widehat m_{1}(z)&=& m_{1}(z)\\
\widehat m_{2}(z)&=& \frac{1}{K}(m_{2}(z)-m_{1}(z)^2)\\
m_{j}(z)&=&
\frac{\sqrt{N}}{2\sigma\sqrt{\pi z T}}
\int\limits_{0}^{F}
\frac{1}{V}
\left(\frac{F-V}{F}\right)^{j}
\nonumber\\&&\times
\exph{-\frac{N(\ln(V/V_{0})-(\mu-\sigma^{2}/2)T)^{2}}{4zT\sigma^{2}}}\rd V
\label{eq:cr:simpleapprox}
\end{eqnarray}
Here $V$ is a scalar and we only have to calculate a single integral over $V$.
After inserting this into Eq.~(\ref{eq:cr:simpelapprox}), we can calculate the loss distribution for a homogeneous portfolio in the second order approximation. 

\subsection{Improved approximation for a homogeneous portfolio}
The second order approach can be improved by approximating the individual terms of the loss distribution instead of approximating the expression as a whole, similar as discussed in \cite{schaefer07}. In case of a homogeneous portfolio the combinatorial sum in Eq. (\ref{eq:cr:combsum}) reduces to
\begin{eqnarray}
\left\langle p^{\rm (loss)}(L)\right\rangle
&=&
\frac{1}{2\pi\Gamma(N/2)}
\int\limits_{0}^{\infty}
\rd z\ 
z^{\left(\frac{N}{2}-1\right)}
\exph{-z}
\int\limits_{-\infty}^{+\infty} \rd\nu \exph{-i\nu L}
\nonumber\\&&\times
\sum\limits_{j=0}^{K}
{K \choose j}
\left(
r^{(\mathrm{D})}(\nu,z)
\right)^{j}
\left(
r^{(\mathrm{ND})}(z)
\right)^{K-j}
\label{eq:cr:improvapp}
\end{eqnarray}
with the non-default term $\left(r^{(\mathrm{ND})}\right)^{K-j}$ where
\begin{eqnarray}
r^{(\mathrm{ND})}&=&
\int\limits_{F}^{\infty} \rd V
\frac{\sqrt{N}}{2\sigma V\sqrt{\pi z T}}
\nonumber\\&&\times
\exph{-\frac{N(\ln(V/V_{0})-(\mu-\sigma^{2}/2)T) ^{2}}{4zT\sigma^{2}}}
\\&=&
\frac{1}{2}+\frac{1}{2}\mathrm{Erf}
\left(\frac{\sqrt{N}(\ln(F/V_{0})-(\mu-\sigma^{2}/2)T) }{2 \sigma \sqrt{zT}}
\right)
\label{eq:cr:nondefault}
\end{eqnarray}
and the default term $\left(r^{(\mathrm{D})}(\nu,z)\right)^{j}$ where
\begin{eqnarray}
r^{(\mathrm{D})}(\nu,z)&=&
\int\limits_{0}^{F} \rd V
\exph{\frac{i\nu}{K}\frac{F-V}{F}}
\nonumber\\&&\times
\frac{\sqrt{N}}{2\sigma V\sqrt{\pi z T}}
\exph{-\frac{N(\ln(V/V_{0})-(\mu-\sigma^{2}/2)T) ^{2}}{4zT\sigma^{2}}} 
\end{eqnarray}
In the homogeneous case $V$ is a scalar variable. The approximation follows the same principles as in the previous section, resulting in
\begin{eqnarray}
&&
\int\limits_{-\infty}^{+\infty} \rd\nu \exph{-i\nu L}
\left(r^{(\mathrm{D})}(\nu,z)\right)^{j}
\nonumber\\&=&
\int\limits_{-\infty}^{+\infty} \rd\nu
\exph{
i \nu 
\left(
\frac{j}{K}  m_{1}(z) -L\right)
-\frac{\nu^2 j}{2K^{2}}
\left(m_{2}(z)-m_{1}(z)^2\right)
}
\\&=&
\sqrt{\frac{2\pi K^{2}}{j\left(m_{2}(z)-m_{1}(z)^2\right)}} \exph{-\frac{(LK-jm_{1}(z))^2}{2j \left(m_{2}(z)-m_{1}(z)^2\right)}}
\end{eqnarray}
In this approximation, the non-default terms given by Eq. (\ref{eq:cr:nondefault}) can be calculated exactly. They correspond to a delta peak at $L=0$. Another advantage over the approach presented in Eq. (\ref{eq:cr:simpelapprox}) is that the approximation is performed for each number of defaults $j$ separately and weighted by $j/K$ accordingly. 
Here, the omitted third term is of the order $j/K^{3}$ and thereby much smaller than the third term of the simple second order approximation (\ref{eq:cr:simpleapprox}), which would be of the order $1/K^{2}$.  Thus, when approximating each term in the combinatorial sum separately, we obtain an improved result. Insertion into (\ref{eq:cr:improvapp}) leads to\begin{eqnarray}
\left\langle p^{\rm (loss)}(L)\right\rangle
& \approx &
\frac{1}{2\pi\Gamma(N/2)}
\sum\limits_{j=0}^{K}
{K \choose j}
\int\limits_{0}^{\infty}
\rd z\ 
z^{\left(\frac{N}{2}-1\right)}
\exph{-z}
\nonumber\\
&& \times
\sqrt{\frac{2\pi K^{2}}{j\left(m_{2}(z)-m_{1}(z)^2\right)}} \exph{-\frac{(LK-jm_{1}(z))^2}{2j \left(m_{2}(z)-m_{1}(z)^2\right)}}
\nonumber\\
&&\times
\left(
\frac{1}{2}+\frac{1}{2}\mathrm{Erf}
\left[\frac{\ln(F/V_{0})-(\mu-\sigma^{2}/2)T }{2 \sigma \sqrt{zT}}
\right]
\right)^{K-j}
\end{eqnarray}
which is the final result.

\section{Application}
\label{s:cr:application}
We now apply the analytically developed model to a specific example. To analyze the impact of correlations, we calculate the loss distribution for different homogeneous portfolios with sizes $K=10$, $K=50$ and $K=100$ with the parameters $V_{0}=100$, $\mu=0.05$, $\sigma=0.15$, $f=75$ and $T=1$.
\begin{figure}[tb]
\begin{center}
\includegraphics[width=0.6\textwidth]{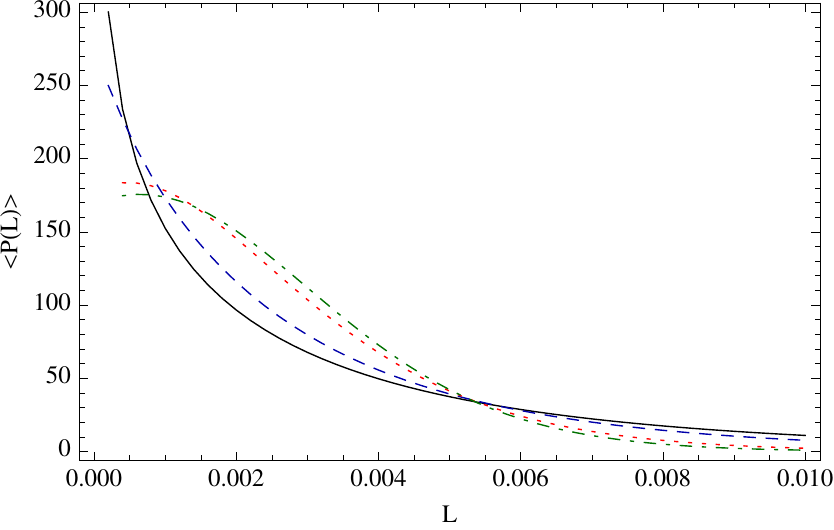}
\caption[The loss distribution for different amounts of randomness in the correlation matrix]{The loss distribution for $K=10$, $\sigma=0.15$, $\mu=0.05$, $T=1$, $V_{0}=100$, $f=75$ and different amounts of randomness in the correlation matrix, $N=K$ (solid black), $N=2K$ (dashed blue), $N=10K$ (dotted red), $N=30K$ (dot-dashed green).}
\label{fig:eq:cr:convergence}
\end{center}
\end{figure}
\begin{figure}[tp]
\begin{center}
\subfloat[K=10]
{
\includegraphics[width=0.5\textwidth]{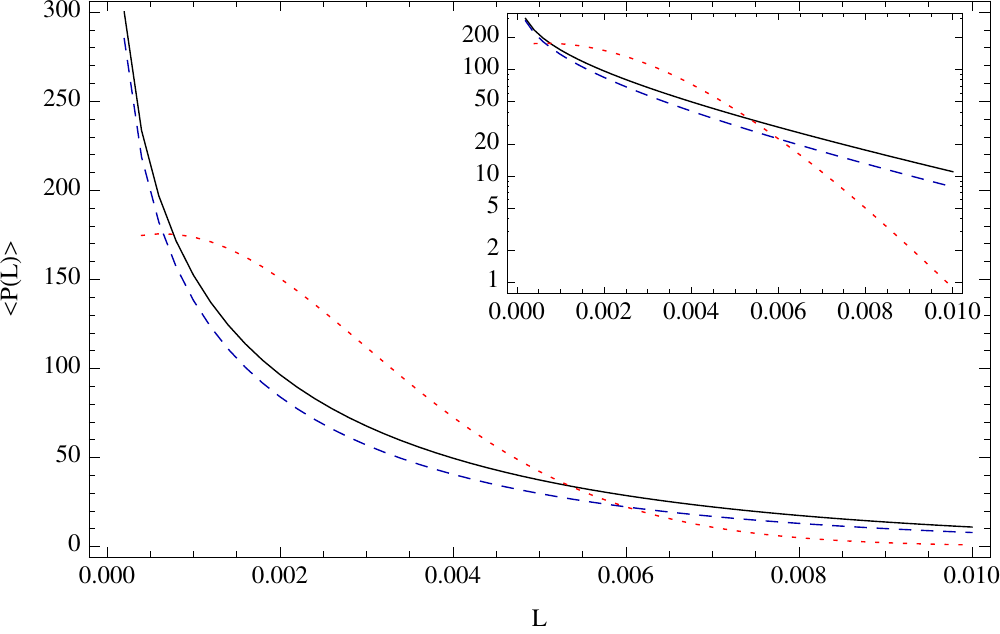}
}\\
\subfloat[K=50]
{
\includegraphics[width=0.5\textwidth]{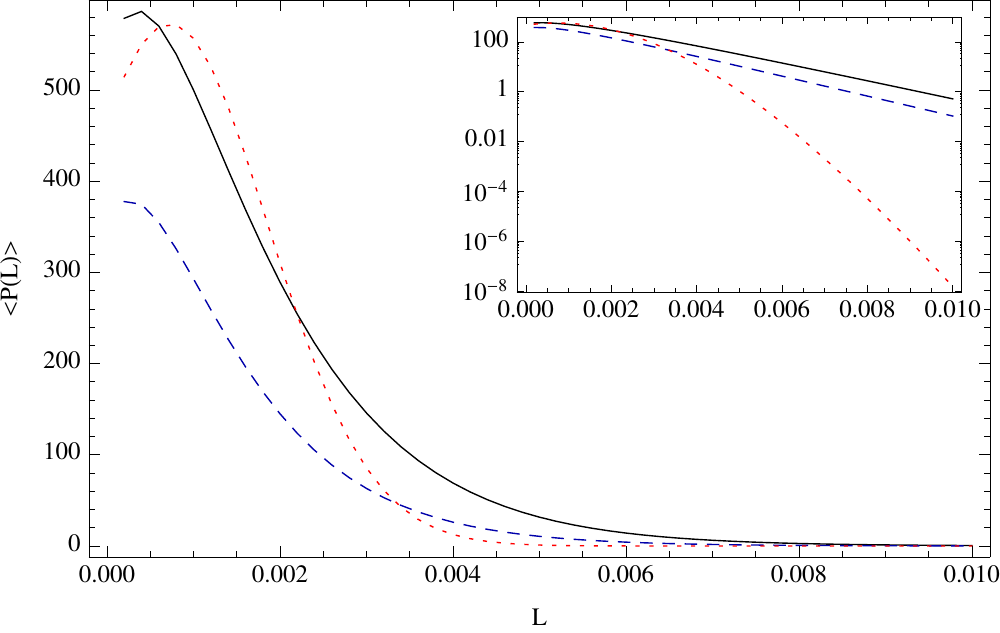}
}\\
\subfloat[K=100]
{
\includegraphics[width=0.5\textwidth]{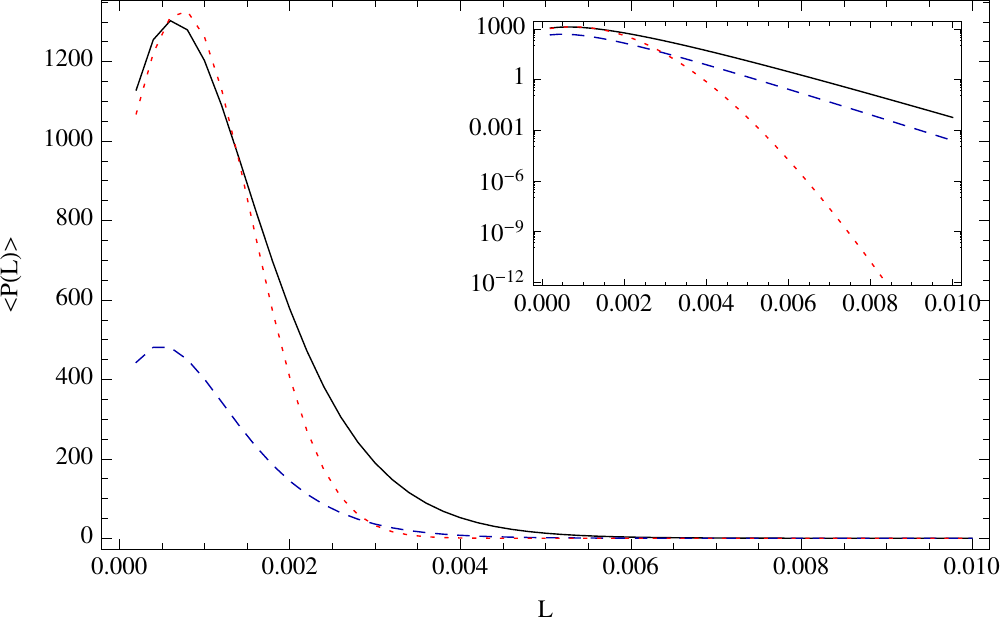}
}
\caption[The loss distribution homogeneous portfolios of different sizes]{The loss distribution of a homogeneous portfolio with $\sigma=0.15$, $\mu=0.05$, $T=1$, $V_{0}=100$, $f=75$ and different values of $K$. The blue dashed line represents the simple approximation; the solid black line represents the improved approximation. Both have been calculated for the strongest random correlations, $N=K$. The uncorrelated case is given by the red dotted line,  calculated with the improved approximation with $N=30K$.}
\label{fig:cr:results}
\end{center}
\end{figure}
As stated in the previous section, we can control the amount of correlation in our model with the parameter $N$.
Since we only consider correlation matrices with full rank, we obtain the strongest correlations if we choose $N=K$.
For $N\to\infty$, the correlation matrix becomes the unit matrix. Thus, this represents the transition to a system without correlations. As we have to evaluate the loss distributions numerically, the limit $N\to\infty$ has to be properly interpreted. We need to identify a value for which this convergence is valid in good approximation. 
Figure \ref{fig:eq:cr:convergence} illustrates the loss distribution for $K=10$ and different values of $N$. Our study indicates that a value of $N=30K$ is a good choice for approximating the uncorrelated case and is still numerically feasible. 
The results are presented in Fig. \ref{fig:cr:results}. For all portfolio sizes, $K=10$, $K=50$ and $K=100$, we obtain heavier tails of the loss distribution of the correlated portfolio compared to the uncorrelated case. Even the simple approximation, represented by the dashed blue curve, exhibits these heavy tails. With the inserted logarithmic plots, we can identify a nearly power-law decay of the loss distribution for the correlated case. 

The distributions become narrower for larger values of $K$. However, the tails of the correlated case remain heavier than those of the uncorrelated case. While both approximations yield similar results for $K=10$, their difference becomes larger with $K$. As both approximations have to be performed numerically, the improved approximation is always favored. However, the tail behavior remains the same, even for the simple approximation, as indicated by the logarithmic scaled inserts in Fig. \ref{fig:cr:results}.  This is a strong indication that the tails of the loss distribution are vastly underestimated if correlations are not taken into account.

Due to the approximation, the normalization of the loss distribution is not exact. Especially the normalization of the simple approximation is problematic for large values of $K$. The normalization might also be used as an indication for the quality of the approximation. The improved approximation exhibits a delta peak at $L=0$, as the non-default terms can be calculated exactly. However, the interval $[0;0.0002[$ was not evaluated due to numerical feasibility.

In our example, we do not vary the maturity time $T$, ie, we choose $T=1$. One can increase $T$ to estimate the evolution of the loss distribution. However, this evolution depends strongly on the drifts $\mu_{k}$ and standard deviations $\sigma_{k}$. Depending on their value, the exposure to default risk can either increase or decrease.

\section{Conclusions}
\label{s:cr:summary}

To assess the risk of a credit portfolio, it is crucial to take correlations between obligors into account. We consider the Merton model, in which defaults and recoveries are determined by the underlying asset processes. The correlation matrix of the asset returns has to be estimated from historical time series. This is not always easy, because the correlations change in time, ie,
they are non-stationary.
Since only time series up to a certain length can be used, the correlation coefficients contain a specific type of randomness, see \cite{laloux99}. Several methods have been put forward to estimate
and to reduce this ``noise''. Thus, we assume that such a noise reduction has been done. The
corresponding ``true'' correlation coefficients and matrices are the proper input
for the structural credit risk model of the Merton type that we consider.
We discussed this issue of noise reduction to emphasize that the random matrix approach in that context is based on a very different motivation as compared to the present application.

Searching for generic properties, we devised the present random matrix approach. Instead of calculating the portfolio loss distribution for a specific correlation matrix, we average over an ensemble of random correlation matrices. Our approach transfers concepts of statistical physics. In quantum chaos,
the average over an individual, long spectrum equals the average over an ensemble of random
matrices, if the level number is very high. We expect that a similar self-averaging property also holds
here. This line of reasoning is supported by the following consideration: The correlation coefficients are varying functions in time, because the business relations of the companies change.  This implies that a correlation matrix over a somewhat longer period in time is a varying quantity, ie, it corresponds to some kind of ensemble.

In our model the average correlation level is zero and we assume that there is no branch structure in the correlations. The fluctuation strength of individual correlations is controlled by a single parameter.
This ansatz allowed us to estimate generic statistical properties of the Merton model.
Some features are not taken into account which are present in empirical data, such as jumps or an overall positive correlation level. Those features are difficult to treat completely analytically. However, even in our simple setup we obtain a heavy--tailed loss distribution.
In this sense our model can be used to estimate a lower bound of the risk embedded in a credit portfolio.

Our results clearly demonstrate that the risk in a credit portfolio is heavily underestimated if correlations are not taken into account. 
Even for random correlations with an average correlation level of zero, we observe very slowly decaying  portfolio loss distributions. In contrast, the probability of large losses in uncorrelated portfolios is significantly reduced within the Merton model.

The results are especially relevant for CDOs, bundles of credits that are traded on equity markets. CDOs are constructed in order to lower the overall risk. The components of a CDO can be exposed to large risks. It is often believed that the CDO has a significantly lower risk. We showed that this diversification only works well if the correlations in the credit portfolio are identical to zero. 

\section*{Acknowledgment}
M.C.M. acknowledges financial support from Studienstiftung des deutschen Volkes.
\section*{Bibliography}
\bibliography{../../bib/literature}
\bibliography{manuscript_arxiv.bbl}
\newpage

\begin{appendix}
\section{Identifying a Gaussian integral in average the price distribution}
\label{app:aveprice}
\begin{eqnarray}
\langle p^{\rm (mv)}(V) \rangle
&=& \int p^{\rm (corr)}(W) p^{\rm (mv)}(V, S W W\tp S) \rd [W] \\
&=& \left(\frac{1}{\sqrt{2\pi}}\right)^{2K+KN}\sqrt{N}^{KN}
\int \exph{-\frac{N}{2}\tr WW\tp} 
\nonumber\\&&\times
\int \exph{-i \omega\tp V}
\exph{-\frac{T}{2}\omega\tp S W W\tp S\omega}
\rd [\omega]
\rd [W]\\
&=& \left(\frac{1}{\sqrt{2\pi}}\right)^{2K+KN}\sqrt{N}^{KN}
\int \exph{-i \omega\tp V}
\nonumber\\&&\times
\int \exph{-\frac{N}{2}\tr W W\tp} 
\nonumber\\&&\times
\exp\Big(-{\frac{T}{2}
\omega\tp S W W\tp S\omega
} \Big)
\rd [W]
\rd [\omega]
\end{eqnarray}
In the last steps, we took advantage of the fact that the term $\omega\tp S W W\tp S\omega$ is a scalar, which can evidently be written as trace. As the trace is invariant under cyclic permutation, we can express this term as $\tr(W W\tp S\omega\omega\tp S)$. Hence, we write
\begin{eqnarray}
\langle p^{\rm (mv)}(V) \rangle
&=& \left(\frac{1}{\sqrt{2\pi}}\right)^{2K+KN}\sqrt{N}^{KN}
\int \exph{-i \omega\tp V}
\nonumber\\&&\times
\int \exph{-\frac{1}{2}\tr(W W\tp (N I+ T S\omega \omega\tp S) )}
\rd [W]
\rd [\omega]\\
&=& \left(\frac{1}{\sqrt{2\pi}}\right)^{2K+KN}\sqrt{N}^{KN}
\int \exph{-i \omega\tp V}
\nonumber\\&&\times
\int \exph{-\frac{1}{2}\sum\limits_{N=1}^{N}(
W_{n}\tp 
(N I+ T S\omega \omega\tp S)
W_{n}
)}
\rd [w_{n}]
\rd [\omega]\\
&=& \left(\frac{1}{\sqrt{2\pi}}\right)^{2K+KN}\sqrt{N}^{KN}
\int \exph{-i \omega\tp V}
\nonumber\\&&\times
\Bigg(
\int
\exp\bigg(
{-\frac{1}{2}
w\tp 
(N I+ T S\omega \omega\tp S)
w
)}
\bigg)
\rd [w]
\Bigg)^{N}
\rd [\omega]
\end{eqnarray}
where $I$ is the unit matrix.
The last step can be accomplished, as the components of $W$ are independent identically distributed, hence we can denote  the $n$-th column vector of $W$, $w_{n}$ by $w$. Thus, we can simplify the integration over the matrix $W$ to the integration over the vector $w \in \mathbb{R}^{K}$ to the power of $N$.
The integral over $\rd [w]$ is simply a Gaussian integral, as indicated by $x_{i}$, $x_{j}$ and $A_{ij}$. As $w_{n}$ consists of $K$ components, this gives an additional factor $\sqrt{2\pi}^{KN}$ and thus leads to
\begin{eqnarray}
\langle p^{\rm (mv)}(V) \rangle&=& \frac{\sqrt{N}^{NK}}{(2\pi)^K}
\int \exph{-i \omega\tp V}
\frac{1}{\sqrt{\mathrm{det}(N I + T S\omega \omega\tp S)}^{N}} 
\rd [\omega]
\end{eqnarray}

\section{Derivation of component-wise average price distribution}
\label{app:aveprice2}
\begin{eqnarray}
\langle p^{\rm (mv)}(V) \rangle
&=& \frac{1}{(2\pi)^K}
\frac{1}{\Gamma(N/2)}
\int
\exph{-i \omega\tp V}
\nonumber\\&=&\times
\int\limits_{0}^{\infty}
z^{\left(\frac{N}{2}-1\right)}
\exph{-(1+ \frac{T}{N}\omega\tp S S\omega)z}  \rd z\,\rd [\omega]\\
&=& \frac{1}{(2\pi)^K}
\frac{1}{\Gamma(N/2)}
\int\limits_{0}^{\infty}
z^{\left(\frac{N}{2}-1\right)}
\exph{-z}
\nonumber\\&&\times
\int
\exph{-i \omega\tp V}
\exph{- \frac{T}{N}\sum\limits_{k=1}^{K} \sigma_{k}^{2}\omega_{k}^{2} z}  \rd [\omega]\rd z\\
&=& \frac{1}{(2\pi)^K}
\frac{1}{\Gamma(N/2)}
\int\limits_{0}^{\infty}
z^{\left(\frac{N}{2}-1\right)}
\exph{-z}
\nonumber\\&&\times
\prod\limits_{k=1}^{K}
\left[
\int
\exph{-i \omega_{k} V_{k}}
\exph{- \frac{T}{N}\sigma_{k}^{2}\omega_{k}^{2} z}  \rd\omega_{k}
\right]
\rd z\\
&=& \frac{1}{(2\pi)^K}
\frac{1}{\Gamma(N/2)}
\int\limits_{0}^{\infty}
z^{\left(\frac{N}{2}-1\right)}
\exph{-z}
\nonumber\\&&\times
\prod\limits_{k=1}^{K}
\left[
\frac{\sqrt{\pi N}}{\sqrt{zT}\sigma_{k}}
\exph{-\frac{N V_{k}^{2}}{4 Tz\sigma_{k}^{2}}}
\right]
\rd z
\label{eq:cr:meanPValt}\\
&=&
\frac{1}{(2\pi)^K}
\frac{1}{\Gamma(N/2)}
\left(
\prod\limits_{k=1}^{K}
\frac{1}{\sigma_{k}}
\right)
\nonumber\\&&\times
\int\limits_{0}^{\infty}
z^{\left(\frac{N}{2}-1\right)}
\exph{-z}
\left(\frac{\pi N}{zT}\right)^{\frac{K}{2}}
\exph{-\frac{N}{4 Tz}\sum\limits_{k=1}^{K}\frac{V_{k}^{2}}{\sigma_{k}^{2}}} \rd z
\end{eqnarray}

\section{Normalization of the average price distribution}
\label{appnorm}
We consider the variance of the $i$-th price $V_{i}$,
\begin{eqnarray}
\mathrm{var}(V_{i}) = \int \rd [V]\ V_{i}^{2} \langle p^{\rm (mv)}(V) \rangle 
\label{eq:cr:varstart}
\end{eqnarray}
We can solve this integral by using hyperspherical coordinates.
\begin{eqnarray}
\rho \equiv \sqrt{\sum\limits_{k=1}^{K}\frac{V_{k}^{2}}{\sigma_{k}^{2}}} \qquad\mathrm{with}\qquad \frac{V_{i}}{\sigma_{i}} \equiv \rho\, \mathrm{cos}\,\vartheta
\end{eqnarray}
for a chosen component $V_i/\sigma_i$.
Now we can write the integral in Eq. (\ref{eq:cr:varstart}) as
\begin{eqnarray}
\mathrm{var}(V_{i}) &=& \sqrt{\frac{N}{2\pi T}}^{K}
\frac{1}{\Gamma(N/2)}
\left(
\prod\limits_{k=1}^{K}
\frac{1}{\sigma_{k}}
\right)2^{1-\frac{N}{2}}
\int\limits_{0}^{\infty}\rd\rho\, \rho^{K-1}
\nonumber\\&&\times
\int\limits_{0}^{\pi}\rd\vartheta\, \mathrm{sin}^{K-2}(\vartheta) \sigma_{i}^{2}\rho^{2}\mathrm{cos}^{2}(\vartheta)
\left(\sqrt{\frac{N}{T}}\rho\right)^{\frac{N-K}{2}}
\nonumber\\&&\times
\mathcal{K}_{\frac{K-N}{2}}\left(\sqrt{\frac{N}{T}}\rho \right)
\int \rd\Omega_{K-1}
\end{eqnarray}
with the surface of the corresponding $K$-dimensional sphere
\begin{eqnarray}
\int \rd\Omega_{K-1} = \frac{2 \pi^{(K-1)/2}}{\Gamma((K-1)/2)}\prod\limits_{k=1}^{K}\sigma_{k}
\end{eqnarray}
We obtain
\begin{eqnarray}
\mathrm{var}(V_{i}) &=& 2\sigma_{i}^{2}\frac{T}{N}\frac{\Gamma(N/2 + 1)}{\Gamma(N/2)}
=
\sigma_{i}^{2} T 
\end{eqnarray}
Thus, the variance of every $V_{i}$ only depends on the standard deviation $\sigma_{i}$ and the time $T$.

\section{second order approximation of average loss distribution}
\label{app:secorder}
We rearranged the constants so that each term in $r(\nu,z)$ is normalized to unity.
The quantity $r(\nu,z)$ can now be written as
\begin{eqnarray}
r(\nu,z) & = &
\exp\Bigg(\sum\limits_{k=1}^{K}
\ln
\Bigg[
\Bigg(\int\limits_{0}^{F_{k}} \rd V_{k}
\overbrace{
\exph{i\nu f_{k}\left(1-\frac{V_{k}}{F_{k}}\right)}
}^{q\left(\nu, F_{k}\right)}
+
\int\limits_{F_{k}}^{\infty} \rd V_{k}
\Bigg)
\nonumber\\&&\times
\frac{\sqrt{N}}{2\sigma_{k}V_{k}\sqrt{\pi z T}}
\exph{-\frac{N(\ln(V_{k}/V_{k,0})-(\mu-\sigma^{2}/2)T) ^{2}}{4zT\sigma_{k}^{2}}}
\Bigg]
\Bigg)
\label{eq:cr:inhom}
\end{eqnarray}
We expand $q\left(\nu, F_{k}\right)$ as the power series
\begin{eqnarray}
q\left(\nu, F_{k}\right) ={}& \sum\limits_{j=0}^{\infty}\frac{(i \nu f_{k})^{j}}{j!}
\left(\frac{F_k-V_{k}}{F_{k}}\right)^{j}
\label{eq:cr:Q}
\end{eqnarray}
Due to the normalization of $\langle p^{\rm (mv)}(V)\rangle$, after insertion into Eq. (\ref{eq:cr:inhom}) the non-default term and the integral over first term of Eq. (\ref{eq:cr:Q}) become one. Thus, we can start the sum at $j=1$ and obtain
\begin{eqnarray}
r(\nu,z) &=
\exph{\sum\limits_{k=1}^{K}
\ln
\left(
1+\sum\limits_{j=1}^{\infty}\frac{(i \nu f_{k})^{j}}{j!}m_{j,k}(z)
\right)
}
\label{eq:cr:logtodevelop}
\end{eqnarray}
with
\begin{eqnarray}
m_{j,k}(z)&=&
\frac{\sqrt{N}}{2\sigma_{k}\sqrt{\pi z T}}
\int\limits_{0}^{F_{k}}
\frac{1}{V_{k}}
\left(\frac{F_k-V_{k}}{F_{k}}\right)^{j}
\nonumber\\&&\times
\exph{-\frac{N(\ln(V_{k}/V_{k,0})-(\mu-\sigma^{2}/2)T) ^{2}}{4zT\sigma_{k}^{2}}}
\rd V_{k}
\label{eq:cr:MgBm}
\end{eqnarray}
The integrals in Eq.~(\ref{eq:cr:MgBm}) can be expressed with the generalized hypergeometrical function ${}_{p}\mathcal{F}_{q}$. However, the integral representation (\ref{eq:cr:MgBm}) is more intuitive. Moreover, for explicit $m=1,2$ the integrals can be calculated in a closed form, although this results in bulky expressions.

Expanding the logarithm and collecting all terms up to the second order in $f_{k}$ yields
\begin{eqnarray}
\left\langle p^{\rm (loss)}(L)\right\rangle
\approx{}& 
\frac{1}{2\pi\Gamma(N/2)}
\int\limits_{0}^{\infty}
\rd z\ 
z^{\left(\frac{N}{2}-1\right)}
\exph{-z}
\int\limits_{-\infty}^{+\infty} \rd\nu \exph{-i\nu L}
\nonumber\\
&
\times
\exph{\sum\limits_{k=1}^{K}
\left(
i \nu m_{1,k}(z)f_{k}
-\frac{\nu^2 f_{k}^{2}}{2}(m_{2,k}(z)-m_{1,k}(z)^2)
\right)
}
\\&=
\frac{1}{2\pi\Gamma(N/2)}
\int\limits_{0}^{\infty}
\rd z\ 
z^{\left(\frac{N}{2}-1\right)}
\exph{-z}
\int\limits_{-\infty}^{+\infty} \rd\nu
\nonumber\\&\times
\exph{
i \nu 
\left(
\left[\sum\limits_{k=1}^{K} f_{k} m_{1,k}(z)\right] -L\right)
\right.\nonumber\\&\left.
-\frac{\nu^2 }{2}
\left[\sum\limits_{k=1}^{K}f_{k}^{2}\left(m_{2,k}(z)-m_{1,k}(z)^2\right)\right]
}
\end{eqnarray}
Now we can solve the $\nu$ integral leading to Eq.~(\ref{eq:cr:simpelapprox}).

\end{appendix}

\end{document}